# Container late-binding in unprivileged dHTC pilot systems on Kubernetes resources

Igor Sfiligoi

University of California San Diego, isfiligoi@sdsc.edu

Yunjin Zhu

University of California San Diego, yuz247@ucsd.edu

Jaime Frey

University of Wisconsin-Madison, jfrey@cs.wisc.edu

The scientific and research community has benefited greatly from containerized distributed High Throughput Computing (dHTC), both by enabling elastic scaling of user compute workloads to thousands of compute nodes, and by allowing for distributed ownership of compute resources. To effectively and efficiently deal with the dynamic nature of the setup, the most successful implementations use an overlay batch scheduling infrastructure fed by a pilot provisioning system. One fundamental property of these setups is the use of late binding of containerized user workloads. From a resource provider point of view, a compute resource is thus claimed before the user container image is selected. This paper provides a mechanism to implement this late-binding of container images on Kubernetes-managed resources, without requiring any elevated privileges.

## 1 INTRODUCTION

In the current scientific compute landscape, the ownership and location of compute resources is distributed, and the situation is unlikely to change in the near future. Nevertheless, many science users want to utilize more than one homogeneous resource cluster, either due to availability, performance, wait time, or cost reasons.

Compute resources operated by many independent entities are very likely to be heterogeneous, both in terms of hardware capabilities and installed software. To deal with the latter, most dHTC users bring their own software stack as part of their workload, with containerization [1-3] being the most popular solution due to its ease of use and wide infrastructure support.

While manually partitioning a workload among the various resource providers is certainly an option, using an automated system is much more user friendly. This premise is at the root of the distributed High Throughput Computing (dHTC) paradigm, with most successful implementations using an overlay batch scheduling infrastructure fed by a pilot provisioning system [4-5].

The main tenant of pilot-based dHTC is the separation of the compute resource provisioning step and workload tasks' scheduling steps. In particular, the provisioning phase requests compute resources from the resource providers on behalf of a generic service identity, i.e. the *pilot* service. Only after the pilot is granted those resources and the management processes start, an actual user payload is selected, fetched to that compute resource, and executed. Moreover, a single pilot may serve several independent payloads during its lifetime, too. This is often referred to as **late-binding**.

The other important principle of successful dHTC infrastructures is the ability of operating using conventional, **non-privileged** credentials. This minimizes the threat surface of the pilot infrastructure for the compute resource operators, making it an acceptable provisioning mechanism for most of them.

Bare-metal batch computing has long been the preferred method for resource providers to schedule pilot requests. To support containerized payloads, most of them provide either singularity [6-7] or apptainer [8], both of which allow for starting processes inside user-provided container images without requiring elevated privileges. Most of the pilot-based container-aware dHTC implementations have thus been designed around these mechanisms [9]. A summary overview of the architecture is available in Figure 1.

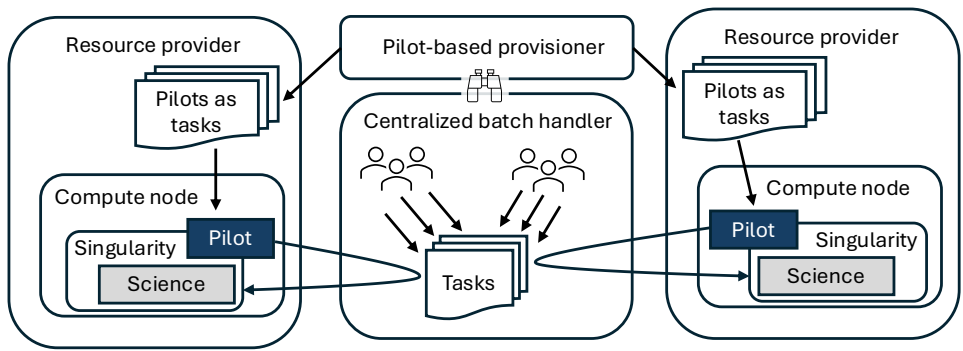

Figure 1: Schematic overview of a traditional container-based pilot-based dHTC system

## 2  THE NESTED CONTAINERIZATION PROBLEM

More recently, Kubernetes has started to become a viable alternative, with the National Research Platform [10] being an example science-focused resource. One defining property of Kubernetes is the requirement that all payloads must be containerized, without exception. This results in the pilot processes themselves running inside a containerized environment.

While unprivileged container runtimes provide many benefits, they typically do not allow for starting of processes in an alternative container image at runtime. I.e., there is typically no support for *nested containerization*.

To work around this limitation, existing dHTC provisioning systems resort to either:

a) **Request limited privileges from the resource providers** for the pilot container [11], in order to implement the nested containerization capabilities. Since not all resource providers are willing to grant such permissions, this limits the breadth of resources that can be made available to the users. It also increases the threat surface of the pilot infrastructure.

b) Avoid the use of nested containerization, by **requesting different resources for user workloads with different container images** [12], and run the pilot processes inside those user-provided container images. This makes the pilot infrastructure significantly harder to maintain in a multi-user environment, since the installed software inside user-provided container images can vary drastically. It also partially breaks the late-binding logic.

Obviously, neither option is truly desirable, as both come with significant drawbacks. In the next section we thus propose an alternative setup that retains full functionality without requiring any privileges not typically granted to regular Kubernetes users.

## 3  USING KUBERNETES MULTI-CONTAINER PODS

In Kubernetes the smallest provisioning request is a ***pod***, which is defined as a set of containers. While a pod can be composed of a single container, there are no drawbacks in requesting a multi-container pod. This paper showcases how one can use multi-container pods to implement a dHTC pilot system that is both unprivileged and supports full late-binding of container images.

### 3.1  Affected pilot functionality

Most traditional pilot-based systems were designed assuming direct execution of payload processes, i.e. the pilot process can directly control the payload process tree. By using independent containers in a Kubernetes pod, this assumption is broken, as the process tree is independently managed by the Kubernetes runtime, instead. This section described the functionality that is affected by this change for one specific, popular pilot runtime, namely HTCondor [13-14]. We believe other pilot runtimes will have very similar requirements.

A traditional HTCondor-based pilot task goes through the following conceptual steps:
  a) An external script validates the working environment, creates the needed HTCondor configuration files, and starts the HTCondor processes
  b) A HTCondor process fetches a payload from a remote task repository, if a matching task is available. Apart from the container image and the desired execution command details, this often includes fetching several associated input files, some of which may be archives that are locally unpacked.
  c) A HTCondor process forks a new process by invoking the container runtime with the user-provided container image to execute the payload startup process, and waits for its completion. User provided files are bind-mounted inside the container environment.
  d) A HTCondor process monitors the forked process tree for both reporting purposes and to ensure that the user payload stays within the permitted constraints. A misbehaving task may be sent a termination signal.
  e) After termination is detected in (c), output files are uploaded to the final destination, and the exit code communicated to the task repository.
  f) The work area is cleaned up and any orphaned processes killed.
  g) Optionally, the pilot may fetch more jobs by looping back to (b).
  h) Finally, the pilot cleans up its own temporary files and terminates.

A summary overview is available in Figure 2.

Apart from the initial setup and final cleanup steps, all other steps involve some interaction between the pilot infrastructure and the user-provided payload. The next few sections describe how this can be achieved in a multi-container Kubernetes pod setup.

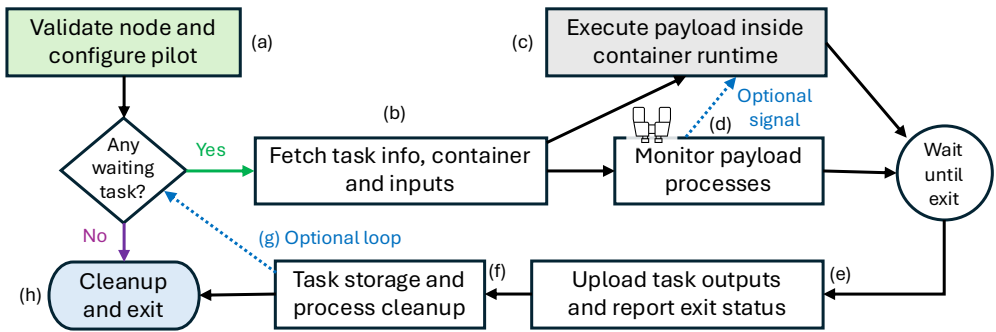

Figure 2: Schematic overview of a pilot life cycle

## 3.2 Sharing of storage

The most basic functionality of a pilot management infrastructure, needed by most of the steps described in the previous section, is the handling of user data, both input and output. There is thus a need for the pilot processes to access the storage area of the user processes. Having a shared storage area is highly desirable to minimize overhead.

Kubernetes pods support multiple storage areas per container [15]. Each such area can be configured independently, with the option of belonging to any subset of the pod containers. It is thus possible to create a pod composed of separate container, one for the pilot infrastructure and one for the user payload processes, that share a common storage area. Both containers have full access to such a shared storage area, providing completely equivalent functionality to the traditional pilot setup.

Furthermore, the pilot container can have an additional private storage area for its own use, fully protecting it from a malicious user payload. A schematic overview of such setup is outlined in Figure 3.

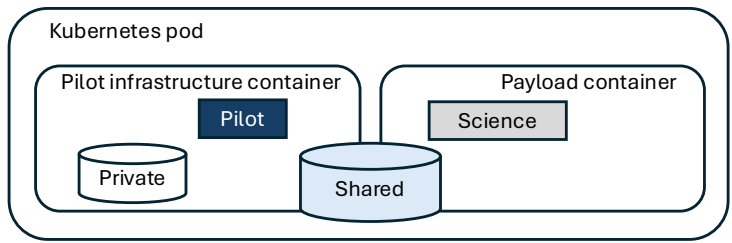

Figure 3: Schematic overview of a multi-container pod

## 3.3 Late-binding of user-provided container images

The major reason for using a multi-container Kubernetes pod setup for the pilot infrastructure was the desire to implement late-binding. While Kubernetes pods must specify an image for each container at creation time, each container can be independently restarted at any time into a different container image [16], without affecting any of the remaining containers. We use this mechanism to replace direct invocation of the container runtime found in traditional pilot systems.

To satisfy Kubernetes requirements, the initial "*payload container*" uses an arbitrary, default container image. The container is also configured to wait for a startup script at a pre-determined path in the shared storage area, to be executed once it becomes available. We implement this waiting logic using simple shell scripting, as we are assuming any reasonable container image will come with a functioning shell executable pre-installed.

The pilot processes start in parallel inside the "*pilot container*". As with traditional pilot setups, they proceed to fetch the user payload, which includes the desired container image as described in step (b) above. At this point the pilot infrastructure issues the Kubernetes command to update the image of the payload container, using an appropriate credential associated with its security context [17]. The only required authorization is the "*pod patch*" role, inside its own namespace, which is typically available to all users (and thus not considered privileged).

Once the payload container has restarted with the updated image, the pilot processes will stage all other necessary files into the shared storage area, and create the payload startup script in the pre-determined path. The user task will thus start inside the properly configured payload container.

The whole sequence is outlined in Figure 4.

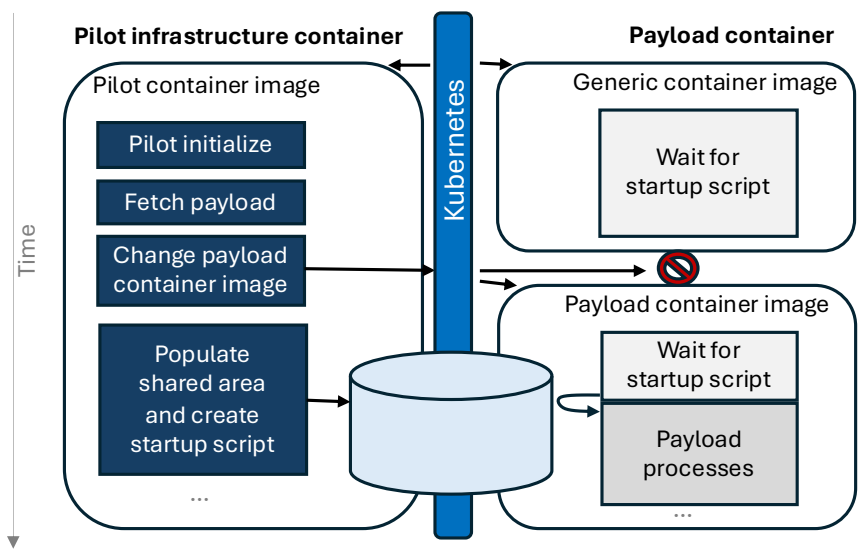

Figure 4: Overview of container late-binding in a multi-container Kubernetes pod

### 3.4 Monitoring and steering the payload processes

As described above, the user-provided payload is executed directly by the Kubernetes infrastructure, in a container that is separate from the one in which the pilot infrastructure runs. There is no direct parent-child relationship between them. The pilot infrastructure thus has to adapt.

Note that by default each container in a pod is completely insulated from the others. This can however be changed by the user [18], exposing a common process tree belonging to all of the containers in a pod. This of course has security implications for the pilot infrastructure, since the payload processes can now see the pilot processes, too. That said, the HTCondor-based pilot infrastructure was always designed with this operational mode in mind, so the pilot infrastructure just has to ensure that the payload processes do not have access to the pseudo-root container UID [19], either at startup or by escalating privileges. Note that the pilot processes should retain the (default) pseudo-root UID, both as a security protection from and to allow for control over the payload processes. These configuration options only affect the pod itself, and are typically available to all users without requiring special privileges.

With the pilot processes able to see the payload process tree, the only remaining difference is the method to determine which processes belong to the payload. By using a well-defined, pre-determined UID for the payload, the pilot infrastructure can assume that all processes belonging to that UID are owned by the payload. Figure 5 provides an example view of the process tree as seen by the pilot infrastructure in such a setup. Note that HTCondor already has an optional mechanism that is conceptually very similar [20].

```
UID        PID  PPID  CMD                                  Pilot container
root         7     0  /bin/sh ./pilot_master                              ⎫ Pilot
root        25     7    /stage/pilot_spawner                              ⎬ processes
root        26     7    /stage/pilot_monitor                              ⎭
root        24     0  /bin/sh                                             ⎤ Startup script
root        46    24    su -c /shared/my_simulation payload               ⎦
payload     47    46      /shared/my_simulation                           ⎫ Payload
payload     49    47        /shared/my_subprocess                         ⎬ processes
65535        1     0  /pause                               Payload container
```

Figure 5: An example view of the process tree as seen by the pilot infrastructure, annotated

With the above adaptations, the pilot infrastructure can both monitor and act on the payload processes like it would have in a traditional setup, retaining full functionality. The same mechanism can be used to determine when the payload processes terminate.

### 3.5 Setting the payload environment and fetching its exit code

The lack of a direct process hierarchy, as outlined in the previous sub-section, prevents the pilot monitoring processes from directly retrieving the exit code of the top payload process. To overcome this limitation, we propose to implement this capability in the payload container's startup wrapper, using the shared filesystem as a means for delivering the necessary information back to the pilot infrastructure.

As described in section 3.3, the payload container already requires a startup wrapper, in order to implement the delay startup logic needed for late binding. Adding exit code checking and file-based reporting to a shell-based script is a routine implementation task. The pilot infrastructure processes will of course need to be modified, too, in order to extract the needed information from the appropriate file.

The same startup wrapper will also be used to properly setup the environment used by the payload processes, by sourcing a configuration file from the same shared filesystem. A schematic overview is available in Figure 6.

Note that, in order to prevent the payload from hijacking the wrapper script or any of its files, we propose to run the startup wrapper script as fake-root inside the payload container, dropping the privileges when forking the top-level payload process itself, just like outlined in Figure 5 above.

### 3.6 Payload cleanup

Using separate containers for pilot and payload processes makes cleanup significantly simpler. By requesting a restart of the payload container, the pilot infrastructure can delegate the process cleanup to the Kubernetes runtime. This operation can be initiated either immediately after existing job termination detection, or delayed until the next job payload is available, simply based on policy.

Having a dedicated shared filesystem also drastically helps. The processes in the pilot container can simply remove all files between each pilot container restart.

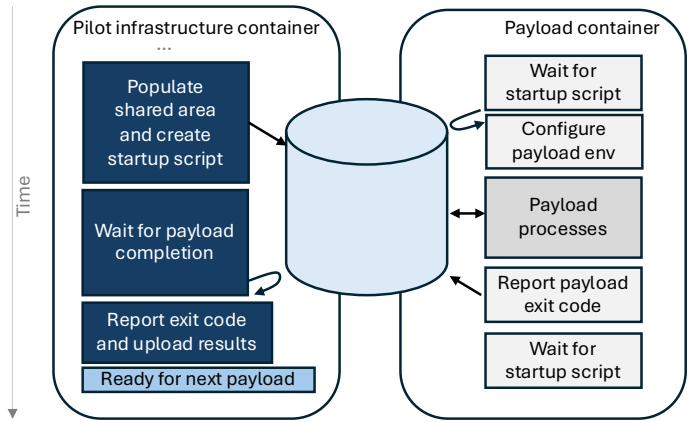

Figure 6: Overview of the setting of payload environment and of the exit code fetching

## 4 PROOF OF CONCEPT EXAMPLES

No full, end-to-end implementation of a multi-container-pod pilot system has been developed at the time of writing of this paper, although we do expect one to become available in the near future.

Nevertheless, we did create a couple proof-of-concept (PoC) examples that exercise the relevant, enabling Kubernetes mechanisms. The two PoCs are both self-contained YAML files, one implementing a fixed sequence and the other using fully dynamic payload fetching, including the container image itself. While toy examples, they do validate the soundness of the proposed approach.

The PoC examples are available in GitHub at `https://github.com/sfiligoi/tnrp-net-tests/tree/master/gil-k8s-pilots-native` [21].

## 5 SUMMARY AND CONCLUSION

Kubernetes is emerging as a popular resource management alternative to traditional batch systems for resource providers in the research computing community. While Kubernetes brings several interesting capabilities, it also presents new challenges for its users. In this paper, we analyze the limitations imposed on pilot-based dHTC systems, which are popular with several scientific communities, and propose an alternative implementation logic that works around those limitations.

The traditional dHTC pilot-based implementations cannot support container-based late-binding without requiring elevated privileges on most Kubernetes setups, due to the implicit need of nested containerization in such setups. The proposed solution avoids the use of nesting, and uses the Kubernetes-native multi-container mechanisms, instead. This paper describes how a pilot infrastructure implementation can be adapted to use these Kubernetes-native mechanisms to its advantage, resulting in a fully functional setup that can we operated without any elevated privileges in most Kubernetes setups.

The focus of this effort was on documenting the necessary steps and providing a proof-of-concept validation of such an approach. We expect to be able to provide a concrete implementation in the near future, to make dHTC pilot infrastructure generally available on Kubernetes-managed resources.

**ACKNOWLEDGMENTS**

This work was funded by the U.S. National Science Foundation (NSF) under grant OAC-2030508. Computing resources used were partially funded by NSF under grant OAC-2112167.

**REFERENCES**

[1] Moreau, D., Wiebels, K. & Boettiger, C. Containers for computational reproducibility. Nat Rev Methods Primers 3, 50 (2023). https://doi.org/10.1038/s43586-023-00236-9

[2] Justin Z Tam, Alexandra Chua, Adyn Gallagher, Denice Omene, Danielle Okun, Dominic DiFranzo, and Brian Y Chen. 2023. A Containerization Framework for Bioinformatics Software to Advance Scalability, Portability, and Maintainability. In Proceedings of the 14th ACM International Conference on Bioinformatics, Computational Biology, and Health Informatics (BCB '23). Association for Computing Machinery, New York, NY, USA, Article 104, 1–5. https://doi.org/10.1145/3584371.3612948

[3] Citation S. Binet and B. Couturier 2015 "docker & HEP: Containerization of applications for development, distribution and preservation"J. Phys.: Conf. Ser. 664 022007, doi: 10.1088/1742-6596/664/2/022007

[4] I. Sfiligoi, D. C. Bradley, B. Holzman, P. Mhashilkar, S. Padhi and F. Wurthwein, "The Pilot Way to Grid Resources Using glideinWMS," 2009 WRI World Congress on Computer Science and Information Engineering, Los Angeles, CA, USA, 2009, pp. 428-432, doi: 10.1109/CSIE.2009.950.

[5] T Maeno et al. 2011 "Overview of ATLAS PanDA Workload Management", J. Phys.: Conf. Ser. 331 072024, doi: 10.1088/1742-6596/331/7/072024

[6] Kurtzer GM, Sochat V, Bauer MW (2017) Singularity: Scientific containers for mobility of compute. PLoS ONE 12(5): e0177459. https://doi.org/10.1371/journal.pone.0177459

[7] Sylabs, Singularity Container Technology and Services, https://sylabs.io, Accessed Jan 22nd, 2025.

[8] Dave Dykstra. 2024. Apptainer Without Setuid, EPJ Web of Conferences 295, 07005, https://doi.org/10.1051/epjconf/202429507005

[9] Namratha Urs, Marco Mambelli, and Dave Dykstra. 2021. Using Pilot Jobs and CernVM File System for Simplified Use of Containers and Software Distribution. In Proceedings of the 30th International Symposium on High-Performance Parallel and Distributed Computing (HPDC '21). Association for Computing Machinery, New York, NY, USA, 255–256. https://doi.org/10.1145/3431379.3464451

[10] Nautilus – National Research Platform. https://nationalresearchplatform.org/nautilus/, Accessed Jan 22nd, 2025

[11] Igor Sfiligoi, Thomas DeFanti, and Frank Würthwein. 2022. Auto-scaling HTCondor pools using Kubernetes compute resources. In Practice and Experience in Advanced Research Computing 2022: Revolutionary: Computing, Connections, You (PEARC '22). Association for Computing Machinery, New York, NY, USA, Article 57, 1–4. https://doi.org/10.1145/3491418.3535123

[12] Fernando Harald Barreiro Megino et al. 2020, "Using Kubernetes as an ATLAS computing site", EPJ Web of Conferences 245, 07025 https://doi.org/10.1051/epjconf/202024507025

[13] Douglas Thain, Todd Tannenbaum, and Miron Livny, 2005. "Distributed Computing in Practice: The Condor Experience" Concurrency and Computation: Practice and Experience, Vol. 17, No. 2-4, pages 323-356. https://doi.org/10.1002/cpe.938

[14] HTCondor Software Suite, https://htcondor.org, Accessed Jan 28th, 2025

[15] Ephemeral Volumes in Kubernetes, https://kubernetes.io/docs/concepts/storage/ephemeral-volumes/, Accessed Jan 28th, 2025

[16] Kubestl set image, https://kubernetes.io/docs/reference/kubectl/generated/kubectl_set/kubectl_set_image/, Accessed Jan 28th, 2025

[17] Configure a Security Context for a Pod or Container, https://kubernetes.io/docs/tasks/configure-pod-container/security-context/, Accessed Jan 28th, 2025

[18] Share Process Namespace between Containers in a Pod, https://kubernetes.io/docs/tasks/configure-pod-container/share-process-namespace/, Accessed Jan 28th, 2025

[19] Configure a Security Context for a Pod or Container, https://kubernetes.io/docs/tasks/configure-pod-container/security-context/, Accessed Jan 28th, 2025

[20] HTCondor Configuration Macros - SLOT<N>_USER, https://htcondor.readthedocs.io/en/latest/admin-manual/configuration-macros.html#SLOT%3CN%3E_USER, Accessed Jan 28th, 2025

[21] Yunjin (Grace) Zhu and Igor Sfiligoi, Example multi-container-pod pilots, https://github.com/sfiligoi/tnrp-net-tests/tree/master/gil-k8s-pilots-native, Accessed Jan 28th, 2025